\documentclass[manuscript]{emulateapj}
\usepackage{natbib}

\newcommand{\lsim}{\mbox{$_<\atop^{\sim}$}}

\newcommand{\lya}{Ly$\alpha$}
\newcommand{\ha}{H$\alpha$}
\newcommand{\hb}{H$\beta$}
\newcommand{\hi}{H~{\small I}}
\newcommand{\hii}{H~{\small II}}

\newcommand{\oiii}{[O~{\small III}]}

\newcommand{\nii}{[N~{\small II]}}

\newcommand{\kms}{km s$^{-1}$}
\newcommand{\lsimeq}{\mbox{$_<\atop^{\sim}$}}

\shortauthors{Scarlata, C. et al.}

\begin{document}

\title{The effect of dust geometry on the \lya\ output of galaxies}

\author{C. Scarlata, J. Colbert, H. I. Teplitz}

\affil{{\it Spitzer} Science Center, California Institute of Technology, 314-6, Pasadena, CA-91125}

\author{N. Panagia\altaffilmark{2,7,9}, M. Hayes\altaffilmark{3}, B. Siana\altaffilmark{4}, A. Rau\altaffilmark{4,5}, P. Francis\altaffilmark{6}, A. Caon\altaffilmark{1,8}, A. Pizzella\altaffilmark{8}, C. Bridge\altaffilmark{1}}

 \altaffiltext{2}{Space Telescope Science Institute, 3700 San Martin Drive, Baltimore, MD 21218, USA}
 \altaffiltext{3}{Observatoire de Geneve, 51, Ch. des Maillettes, CH-1290, Sauverny, Switzerland}
 \altaffiltext{4}{California Institute of Technology, MS 105-24, Pasadena, CA91125}
 \altaffiltext{5}{Max-Planck-Institut f\"ur extraterrestrische Physik, Giessenbachstrasse 1, 85748 Garching, Germany}
 \altaffiltext{6}{Research School of Astronomy and Astrophysics, the Australian National University, Canberra 0200, Australia}
 \altaffiltext{7}{INAF/Osservatorio Astrofisico di Catania, Via S.Sofia 78, I-95123 Catania, Italy}
 \altaffiltext{8}{Department of Astronomy, University of Padova, Vicolo dell'Osservatorio 3, I-35122, Padova, Italy}
 \altaffiltext{9}{Supernova Ltd., Olde Yard Village 131, Northsound Road, Virgin Gorda, British Virgin Islands}

\begin{abstract}
  We present the optical spectroscopic follow-up of 31 $z=0.3$ \lya\
  emitters, previously identified by \citet{deharveng2008}.  We find
  that $17$\% of the \lya\ emitters have line ratios that require the
  hard ionizing continuum produced by an AGN. The uniform dust screen
  geometry traditionally used in studies similar to ours is not able
  to simultaneously reproduce the observed high \lya/\ha\ and \ha/\hb\
  line ratios. We consider different possibilities for the geometry of
  the dust around the emitting sources. We find that also a uniform
  mixture of sources and dust does not reproduce the observed line
  ratios. Instead, these are well reproduced by a clumpy dust
  screen. This more realistic treatment of the geometry results in
  extinction corrected (\lya/\ha)$_C$ values consistent with Case~B
  recombination theory, whereas a uniform dust screen model would
  imply values (\lya/\ha)$_C$ higher than 8.7. Our analysis shows that
  there is no need to invoke “ad-hoc” multi phase media in which the
  \lya\ photons only scatter between the dusty clouds and eventually
  escape.
 
\end{abstract}


\section{Introduction}
In the pre-JWST (James Webb Space Telescope) era we depend strongly on
the observation of \lya\ emission for both spectroscopic confirmation
and, often, the actual discovery of very high-redshift ($z>6$)
galaxies. Furthermore, the evolution with redshift of the clustering
properties and luminosity function of \lya\ emitters could be used to
infer important information about the re-ionization epoch, if the
physics governing the \lya\ escape were well understood. In the local
universe, where the physical properties of galaxies can in principle
be determined in great detail, the number of known \lya\ emitting
galaxies is extremely small \citep[on the order of a dozen objects,
e.g.,][]{giavalisco1996,ostlin2009}.

Numerous theoretical works demonstrate how the observed \lya\ emission
and its equivalent width (EW) depend on different factors including
not only the age of the stellar population, the stellar initial mass
function, the metal and dust content \citep{charlot1993}, but also the
relative geometries of interstellar \hi\ and \hii\ regions and the
kinematics of the neutral gas \citep[e.g.,][]{panagia1973a,
  panagia1973b,neufeld1990, verhamme2006, dijkstra2006,
  laursen2007}. All of these studies point to a preferential
attenuation at resonant frequencies, although it is likely that the
homogeneous interstellar medium (ISM) model is a great
over-simplification.  In contrast to the homogeneous case, a
multiphase ISM in which the dust lies in cold neutral clouds (which
\lya\ photons cannot penetrate), would actually serve to preserve
\lya\ photons, leading to a relative enhancement of the \lya\ EW and
the \lya\ / Balmer line ratios
\citep[e.g.][]{neufeld1991,hansen2006}. This latter mechanism has been
proposed to explain the high \lya\ EW sources observed at high
redshift \citep{finkelstein2008geom} and the high \lya/\ha\ ratio
inferred for low-$z$ dusty \lya\ emitters \citep{atek2008,atek2009}.

Recently \citet{deharveng2008} presented the results from a slitless
spectroscopic survey tuned to identify low-$z$ \lya\ emitters with the
Galaxy Evolution Explorer (GALEX) satellite. The survey covered $\sim$
5.6 deg$^2$ and identified 96 \lya\ emitters with redshifts in the
range $0.2 < z < 0.4$. All galaxies have measured \lya\ EW and line
intensities.  The EW distribution of the $z=0.3$ \lya\ emitters is
similar to that at $z\sim3$, but their fraction among star-forming
galaxies is smaller.

We are conducting a program of spectroscopic followup of the GALEX
\lya\ emitters.  In this letter we present the comparison of the
\lya/\ha\ and \ha/\hb\ ratios for the galaxies observed so far.

\section{Observations and data analysis}
\label{sec:data}
Optical spectra of 31 of the Deharveng et al. $z\sim 0.3$ galaxies
were acquired with the the Double-Beam Spectrograph
\citep[DBSP,][]{dbsp} mounted on the Hale 5m telescope at Palomar
Observatory. The objects were selected to be visible from the Palomar
Observatory (in the following fields GROTH, NGPDWS, and SIRTFFL from
Deharveng et al. 2008), and to have $z\lsimeq0.32$ so that \ha\ would
fall in the covered spectral range. The sample is presented in
Table~\ref{tab:sample} available in the electronic version of the
paper. We used the D55 dichroic to split the light at $\sim 5500$\AA\
into a blue and a red channel. The red spectra were acquired with the
600 lines mm$^{-1}$ grating, and cover the wavelength range between
5800\AA\ and 8500\AA. The blue spectra and a detailed description of
the data will be presented in a forthcoming paper (Scarlata et al., in
preparation). Six objects have redshifted \hb\ (\ha) lines outside the observed
spectral range. The exposure times
ranged
between $2\times 200$s and $4\times 600$s.  During the observations
seeing ranged from 1 to 1\farcs5.  The galaxies were observed using a
1\farcs5 wide slit, providing a spectral resolution of 8\AA\ (i.e.,
280 \kms\ at $z=0.3$). This resolution is high enough to resolve the
\ha\ from \nii\ emission lines. To check for slit loss, we also
observed 8 galaxies with the 5\farcs0 wide slit. 

Bias subtraction, flat-field normalization, wavelength calibration,
and extraction of the 1D spectra were performed with standard IRAF
packages.  The wavelength-dependent instrumental response was removed
by normalizing each spectrum with the response curve derived from the
spectrum of a standard star observed in one of our photometric
nights. The final flux calibration was performed using the total
$i$--band magnitude of each galaxy available from the Sloan Digital
Sky Survey \citep[SDSS,][]{sdss}. We convolved each spectrum with the
filter transmission curve\footnote{The 900\AA\ width of the $i$ filter
  is fully contained within the observed red spectra.} and compare the
resulting flux with the total SDSS flux. With this approach the
calibrated spectra are simultaneously corrected for slit losses, due
to the 1\farcs5 size of the slit\footnote{The median half--light
  radius of the sample galaxies is $\sim$0\farcs9.}. Although valid
for the stellar component (i.e., the continuum) of the galaxies, if
the gas is more concentrated than the stars this correction might
introduce a systematic overestimate of the emission line intensities
(without affecting the line ratios). In order to check this effect, we
computed $f_{1\farcs5}/f_{5\farcs0}$, i.e.  the ratio between the \ha\
fluxes measured from the 1\farcs5 and the 5\farcs0 slit, for those
galaxies for which the two spectra were available. On average, the
line intensities agree to within 25\%. This potential source of error
would affect the ratio between our measured optical line intensities
and that of \lya, in the direction of systematically underestimating
the \lya/\ha\ ratio. One source has a \lya/\ha/ ratio $>5$. Even for
this object a 25\% error on the \ha\ would not result in a \lya/\ha\
ratio greater than the case~B value.

We measured line intensities and EWs by fitting a Gaussian to the
emission lines, and checked the results using the IRAF package SPLOT.
All Balmer emission line intensities were corrected for stellar
absorption, fitting the continuum of each spectrum between 2600 and
6600\AA\ rest--frame with synthetic templates from the Bruzual \&
Charlot library \citep{bruzual1993,bruzual2003}. Templates with
constant star formation history, different age and metallicity were
employed. The median value of the stellar absorption rest frame EW for
\ha\ and \hb\ in our galaxies are, respectively, 3.38\AA\ and 4.18\AA.
For an EW of 20\AA, these values correspond to a flux correction of
about 21\% and 27\% in \ha\ and \hb, respectively.  We also applied a
\lya\ stellar absorption correction of 10\AA\ EW to all galaxies with
a best--fit age older than 35 Myrs, following
\citet{vallsgabaud1993}.

\section{AGN identification}
\begin{figure}
\epsscale{1.2}
\plotone{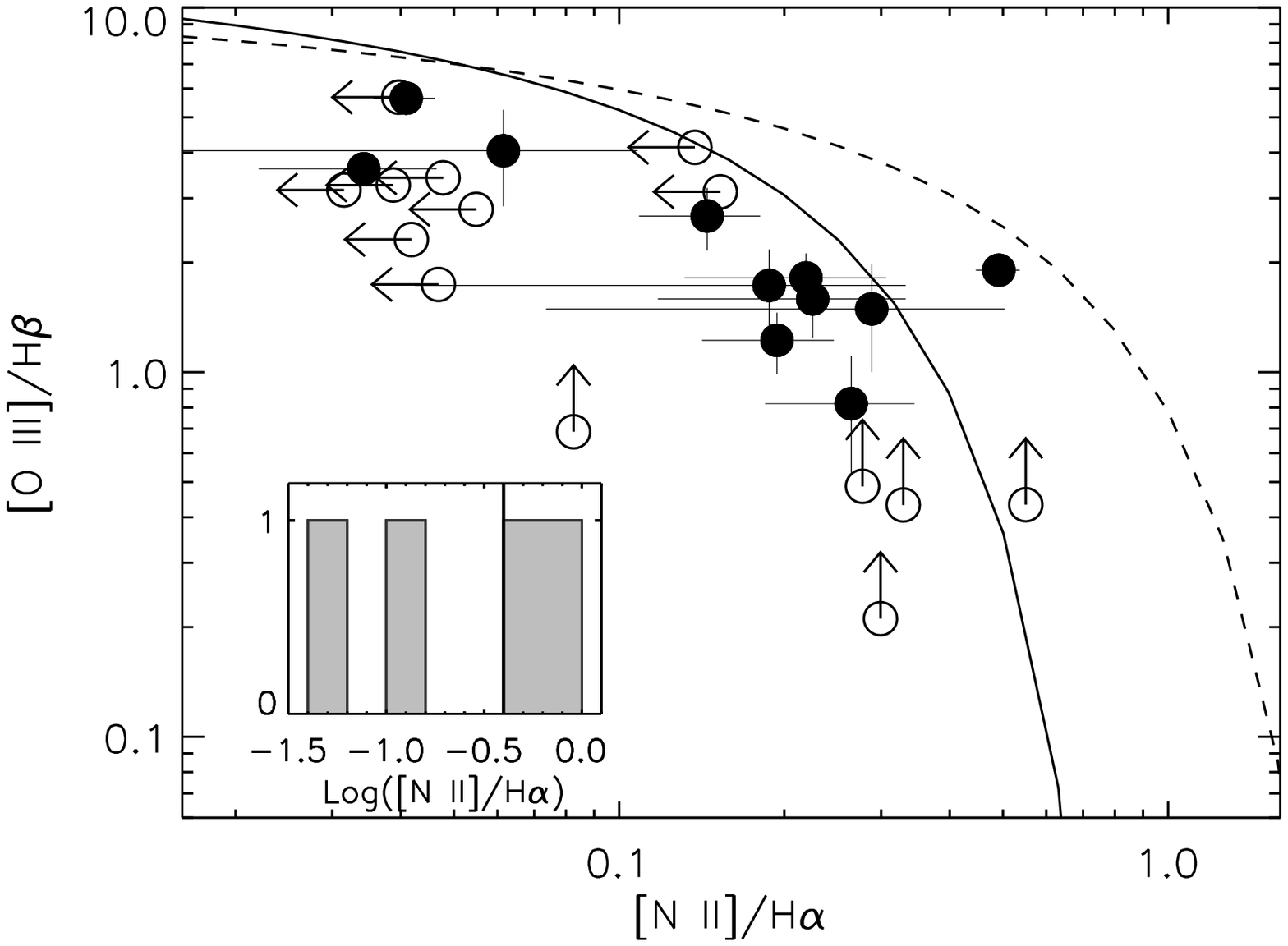}
\caption{\label{fig:agn} BPT diagnostic diagram. \oiii/\hb\ is plotted
  against \nii/\ha\ for the GALEX \lya\ emitters (filled circles and
  limits).  When a line was not detected, we assign the $3\sigma$
  upper limit to the line intensity. The dashed and solid curved lines
  represent the \citet{kewley2001} and \citet{kauffmann2003}
  classification lines, respectively, which divide \hii\ regions
  (below the lines) from the region occupied by AGN (above the
  lines). To be conservative, we use the \citet{kauffmann2003} curve
  to identify AGN in our sample. Inset: histogram of \ha/\nii\ ratio
  for the 4 galaxies for which \hb\ fell outside the wavelength range
  covered by the spectra. A galaxy is classified as an AGN if
  $\log([NII]/H\alpha)>-0.4$. One object, not shown in this figure,
  was classified as an AGN based on the width of the \oiii\ emission
  line (see text).}
\end{figure}

\citet{deharveng2008} identified broad-line AGN in their sample of
\lya\ emitters as those objects with \lya\ line widths broader than
1200 \kms. Narrow line AGN could not be identified since other
diagnostic lines of AGN activity were either too faint, or fell in a
noisy part of the UV spectra.  In order to identify the narrow line
AGN, we have used the classic \citet[][BPT]{baldwin1981} diagram of
\oiii/\hb\ versus \nii/\ha (shown in Figure~\ref{fig:agn}).

The theoretical maximum line ratios possible for pure stellar
photoionization are shown as a dashed curve in Figure~\ref{fig:agn}
\citep{kewley2001}. To be conservative, we use the
\citet{kauffmann2003} empirical separation between active and
star-forming galaxies (solid line). We find that in 2 of our \lya\
emitters the gas is likely to be ionized by an active nucleus. We note
that both cases are in the region of the diagram where stellar
photoionization cannot be excluded on the basis of theoretical
calculations \citep{kewley2001}.  Among the six galaxies that could
not be placed on the BPT diagram, we could still classify 5 of them
using either \nii/\ha\ ratio, or the line width. We identify the AGN
as those galaxies for which $\log \left( [N~II] / H~\alpha \right)
\geq -0.4$ \citep{miller2003,carter2001}.  We add 2 AGN based on the
\nii/\ha\ ratio.  Finally, one of the two galaxies for which the \ha\
is too red to be covered by our spectrum has an \oiii\ line width
typical of an AGN ($\sim 1090$ \kms). To summarize, we identify 5 AGN
out of a sample of 30 galaxies, i.e, the fraction of \lya\ emitters
classified as AGN is 17\%, consistent with the recent measurement by
\citep[][15\%]{cowie2009}, and marginally consistent with the
measurement by \citet[][43$^{+18}_{-26}\%$]{finkelsteinagn}. However,
Finkelstein et al. (2009) classify AGN using also the presence of high
ionization emission lines, that may be indicative of a high-excitation
star forming galaxy, rather than an active nucleus.

In the following analysis we only consider those 20 galaxies that are
classified as star-forming, and with {\it both} \ha\ and \hb\ detected
in the spectrum.

\section{Dust Geometry and reddening}
In the absence of scattering, the EW of a line is by definition
unaffected by dust extinction. For this reason, the \lya\ EW has often
been used to study the effect of scattering on the \lya\ photons, by
comparing the observed values with those expected theoretically
\citep[e.g.,][]{giavalisco1996}.  The intrinsic \lya\ EW is, however,
predicted to vary by more than 2 orders of magnitude within the first
ten Myrs of a star-burst \citep{charlot1993}, and depends on the IMF
of the stellar population. 

The observed line intensities depend only on the number of ionizing
photons and the attenuation produced by dust along the line of
sight. Compared to EW, line ratios are a more direct probe of the
obscuration of the line photons. For this reason, the comparison of
the ratios of resonant to non-resonant lines (i.e., \lya/\ha) with the
ratio of two non-resonant lines (\ha/\hb) is optimal for quantifying
the effects of dust extinction and resonant scattering on the escape of
\lya\ photons.

\begin{figure}
\epsscale{1.2}
\plotone{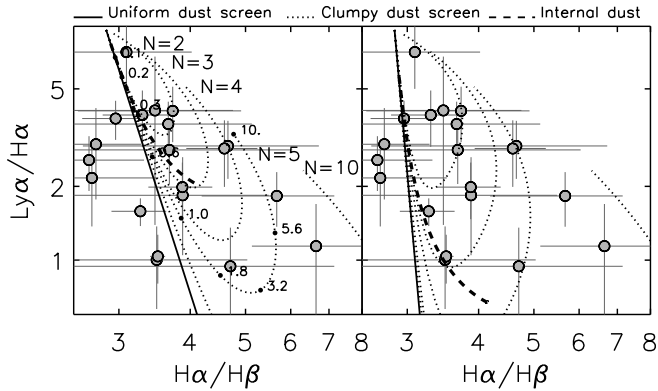}
\caption{\lya/\ha\ ratio against \ha/\hb\ ratio. The various curves
  show the behavior of the line ratios as a function of increasing
  optical depth of the dust, assuming three different geometries for
  the dust/source distribution: uniform dust screen (solid line),
  clumpy distribution (dotted lines) for different values of the
  average numbers of clumps, $N$, along the line of sight, and
  internal dust model (dashed line).  The numbers near the black dots
  on the curve for the clumpy medium with $N=5$ show $\tau_V$ at that
  position. In the left panel all models are computed assuming
  $\tau_{Ly\alpha}=\tau_{1216}$, while in the right panel we account
  for the scattering of the \lya\ photons assuming
  $\tau_{Ly\alpha}=3\times \tau_{1216}$ (see text for
  details). \label{fig:geometry}}
\end{figure}

In Figure~\ref{fig:geometry} we show the \lya/\ha\ ratio as a function
of the observed \ha/\hb\ ratio. Both quantities are corrected for
stellar absorption. In the absence of dust, for Case~B recombination,
$T_e\sim 10^4$ K, and $n_e=10^2$ cm$^{-3}$, the expected line ratios
are 2.86 and 8.7, for \ha/\hb, and \lya/\ha\ respectively
\citep{pengelly1964}. The galaxies show a broad range of \ha/\hb\
values going from the value predicted in Case~B recombination theory,
up to an \ha/\hb$\sim 7$. Apart from one galaxy, the \lya/\ha\ ratio
is always smaller than $50$\% of the value predicted for Case~B.

In order to interpret the observed line ratios, we will first ignore
the effects of resonant scattering on the \lya\ photons, and consider
only the effect of the dust extinction (derived using the \ha/\hb\
ratio) on the \lya/\ha\ ratio. Traditionally, this analysis is carried
out under the simplifying assumption that the dust is distributed
uniformly in a screen far from the source
\citep[e.g.][]{giavalisco1996, hayes2007, atek2008,
  finkelstein2008geom,finkelstein2009}. Although valid for point like
sources, this simple approximation is found not to hold in more
complex systems, as shown by the analysis of Galactic and
extragalactic \hii\ regions, where the uniform dust screen
approximation fails to reproduce the observed Balmer line ratios
\citep[e.g.,][]{natta1984,caplan1986}. More complex geometries do
predict different attenuation laws, with the net effect of removing
the linearity between the dust optical depth and the logarithm of the
observed line intensity.

In order to illustrate this effect, we consider here three different
geometries of the dust around the line emitting regions: $i$) uniform
dust screen in front of the emitting sources; $ii$) dust distributed
in clumps; $iii$) sources uniformly mixed within uniform dust. The
light attenuation\footnote{The attenuation is defined as
  $I_{\lambda}/I^0_{\lambda}$, where $I_{\lambda}$ and $I^0_{\lambda}$
  are the observed and intrinsic intensities at $\lambda$.} has a
different behavior in the three cases. In model $i$) the attenuation
is given by the classical $\exp (-\tau_{\lambda})$, where the optical
depth $\tau_{\lambda}$ follows the appropriate extinction curve, and
depends only on the intrinsic grain properties. We assume that the
dust/gas relative geometry does not affect the physical properties of
the grains, and we adopt the extinction curve as parametrized by
\citet[][$\tau_{\lambda}/\tau_{V}$]{cardelli1989}. In model $ii$)
\citet{natta1984} calculate the attenuation assuming that the dust
clumps are Poissonian distributed, and that all clumps have the same
optical depth, $\tau_{c,\lambda}$. Within each clump,
$\tau_{c,\lambda}$ follows the \citet{cardelli1989} extinction law. If
$N$ is the average number of clumps along the line of sight, then they
find that $I_{\lambda}/I^0_{\lambda}= \exp [-N\,(1-
e^{-\tau_{c,\lambda}})]$. For a large average number of clumps, and a
moderate value of $\tau_V$ for each clump, the attenuation tends to
the uniform dust screen. However, the two models differ greatly if the
clumps have large optical depth. In the limit of opaque clumps, the
attenuation tends towards the constant value of $e^{-N}$, i.e., the
extinction becomes independent of wavelength, because only the
un-extinguished light can be seen. In the third case considered,
$I_{\lambda}/I^0_{\lambda}$ goes as $(1 -
e^{-\tau_{\lambda}})/\tau_{\lambda}$ \citep{mathis1972}. In this model
for $\tau_{\lambda} \rightarrow \infty$, the attenuation reaches the
asymptotic value of $ \tau_{\lambda}^{-1}$. Curves corresponding to
the three models are shown in Figure~\ref{fig:geometry} with different
line styles as indicated on top of the figure. We use $\tau_V$ to
parametrize the observed line ratios in each of the three models.  The
black dots along the curve of the clumpy distribution with $N=5$ are
labeled with the optical depth $\tau_V$ at that position.

In reality, \lya\ photons will be scattered within the clumps,
effectively enhancing their chances of being absorbed by dust. This
effect can be folded in into the model by assuming that the optical
depth for \lya\ photons, $\tau_{Ly\alpha}$, is higher than the
corresponding continuum optical depth, i.e., $\tau_{Ly\alpha}=s\times
\tau_{1216}$. The left panel of Figure~\ref{fig:geometry} shows the
same models described above, but computed accounting for the
scattering of the \lya\ photons, assuming, as an example, $s=3$. As
expected, the net effect of increasing the optical depth of the \lya\
photons results in lower values of the \lya/\ha\ ratios for small
optical depth.  However, for high values of the optical depth, the
asymptotic behavior of the curves remains unchanged, and so do our
conclusions.

\section{Discussion}
From Figure~\ref{fig:geometry} we can see that at \ha/\hb\ $\sim 4$
the uniform dust screen model (solid line) predicts an observed
\lya/\ha\ ratio of \lsim 10\% of the Case~B value. Clearly, the
observed values of \lya/\ha\ for \ha/\hb$\ge 4$ are much higher than
the prediction of a simple dust screen. Correcting the \lya\
luminosities using the attenuation derived for a uniform dust screen
would result in a corrected line ratio --hereafter (\lya/\ha)$_C$--
well above the predicted value of 8.7.  \citet{neufeld1991} and
\citet{hansen2006} discuss a multi-phase geometry, where the dust is
confined in cold clumps and immersed in ionized gas.  A topology of
this kind could serve to shield the \lya\ photons from the effect of
the dust absorption, because resonant photons would be scattered {\it
  on the surface} of the clumps, without crossing them. The Balmer and
continuum photons, not affected by the presence of the neutral
Hydrogen, would pass through the clumps, leading to a relative
enhancement of the \lya\ EW {\it and} \lya\ / Balmer line
ratios. Given the stochastic nature of the \lya\ scattering, this
model should also result in some objects with uncorrected \lya/\ha\
greater than the predicted Case~B value. As noted above, we do not
observe any galaxy with \lya/\ha$>8.7$. Furthermore, if the continuum
is affected by the dust while the \lya\ photons are not, then this
should result in a correlation between the \lya\ EW and the observed
\ha/\hb\ ratio. In fact the four most extreme cases (with \lya/\ha\
ratios more than 10 times higher the dust screen value) have among the
lowest EW.

A model with the dust uniformly mixed with the sources is not able to
reproduce the galaxies with high line ratios (dashed line in
Figure~\ref{fig:geometry}). This is due to the asymptotic behavior of
the attenuation curve for large optical depth. While in a uniform dust
screen the \ha/\hb\ (and \lya/\ha) can reach arbitrarily high (low)
values, in the uniformly mixed model the observed \ha/\hb\ ratio
reaches the asymptotic value of $2.86\times \tau({\rm
  H\beta})/\tau({\rm H\alpha})=4.15$, and similarly a \lya/\ha\ of
$8.7\times \tau({\rm H\alpha})/\tau({\rm Ly\alpha})=2.1$. In this
model the light that is able to reach the observer must come through a
thin superficial shell of the galaxy.

Figure~\ref{fig:geometry} demonstrates that the clumpy dust screen can
explain the observed data. By varying the average number of clumps
along the line of sight, and the intrinsic optical depth of the
clumps, the clumpy distribution of dust allows \ha/\hb\ ratios as high
as 6 while still allowing a \lya/\ha\ ratio of $\sim 2$. This is due
to the fact that for high values of $\tau_V$ only the fraction of
light that is traveling along sight lines relatively clear of clumps is
able to escape the galaxy.  We consider an average number of clumps
$N$ along the line of sight in the range from 2 to 10. It is
reasonable to expect that galaxies may have different physical
structure, depending on the orientation and size of the galaxies, and
that they can be statistically described by different values of $N$.

For three galaxies \ha/\hb\ ratio is consistent with 2.86 (case~B),
while the \lya/\ha\ ratios are $\sim 70$\% lower than Case~B would
predict. Such galaxies require an additional mechanism to reduce \lya\
without affecting the Balmer photons. Possibly, the \lya\ photons pass
through a neutral Hydrogen cloud, without scattering on its
surface. This would not alter the observed Balmer line ratios but
could suppress part of the \lya\ photons. Kinematically resolved
spectra might test this hypothesis, depending on the kinematics of the
\hi\ with respect to the \hii\ region.

\section{Conclusions}
We have presented the analysis of optical spectra of a sample of 31
$z\sim 0.3$ \lya\ emitters, previously identified by
\citet{deharveng2008}.  We find that in 17\% of the \lya\ emitters the
line ratios require the hard ionizing continuum produced by an
AGN. This fraction is in agreement with the fraction of AGN discovered
in local samples of emission line galaxies, covering a similar range
of luminosities.

A uniform dust screen is not able to simultaneously reproduce the
observed high values of \lya/\ha\ and \ha/\hb\ line
ratios. Traditionally, these are explained with {\it ad hoc}
multiphase models in which \lya\ photons are shielded from the dust,
while non-resonant photons are not \citep[e.g.,][]{neufeld1991}. We
consider alternative possibilities for the geometry of the dust around
the emitting sources and find that the observed line ratios are well
reproduced by a clumpy dust distribution. We cannot rule out the
possibility that a combination of dusty and uniform medium is
responsible for the observed line ratios.  Our analysis shows that
there is no need to invoke models requiring a significant percentage
of \lya\ photons to escape from the galaxy through resonant scattering
that avoids the dust. Furthermore, the assumption of a uniform dust
screen would result in an over-correction of the \lya\ intensity.

\begin{table}[h]
\caption{The sample.\label{tab:sample}} 
\centering
\begin{tabular}{lccccccl}
\hline\hline
Object    &  RA   & DEC    &\lya\ flux\tablenotemark{(a)}    &  \ha\ flux   & \hb\ flux    & $z$\tablenotemark{(b)}    &  Class\tablenotemark{(c)}   \\
   &  (deg)   & (deg)    &erg cm$^{-2}$ s$^{-1}$    &  erg cm$^{-2}$ s$^{-1}$   & erg cm$^{-2}$ s$^{-1}$    &    &     \\
\hline
GROTH--13305 & 215.8867 &  52.6237& 1.70$\times 10^{-15}$ & (1.49$\pm$0.10)$\times 10^{-15}$  &  (2.2$\pm$ 0.1)$\times 10^{-16}$ &  0.2836 &  SF  \\
GROTH--14069 & 215.3526 &  52.6555& 1.80$\times 10^{-15}$ & (2.58$\pm$0.10)$\times 10^{-15}$   &  (7.37 $\pm$0.69)$\times 10^{-16}$ &  0.2582 &  SF  \\
GROTH--17005 & 215.1805 &  52.7188& 2.87$\times 10^{-15}$ & (1.18$\pm$0.08)$\times 10^{-15}$   &  (3.23 $\pm$0.52)$\times 10^{-16}$ &  0.2467 &  SF  \\
GROTH--17525 & 215.8241 &  52.7135& 2.62$\times 10^{-15}$ & (2.58$\pm$0.13)$\times 10^{-15}$   & $<$6.15$\times 10^{-16}$  &  0.2770 &  SF  \\
GROTH--17867 & 215.8429 &  52.7425& 1.21$\times 10^{-15}$ & (1.28$\pm$0.10)$\times 10^{-15}$   &  (2.73 $\pm$0.43)$\times 10^{-16}$ &  0.2774 &  SF \\ 
GROTH--19002 & 214.4387 &  52.7719& 2.94$\times 10^{-15}$ & (5.77$\pm$0.58)$\times 10^{-16}$   & $<$3.04$\times 10^{-16}$  &  0.2439 &  SF  \\
GROTH--36336 & 214.5818 &  53.3393& 2.24$\times 10^{-15}$ & (1.95$\pm$0.15)$\times 10^{-15}$   & $<$1.04$\times 10^{-15}$  &  0.2647 &  SF  \\
GROTH--36896 & 214.9730 &  53.3764& 3.29$\times 10^{-15}$ & (2.53$\pm$0.05)$\times 10^{-15}$   &  (4.46 $\pm$0.50)$\times 10^{-16}$ &  0.1942 &  SF  \\
GROTH--5715  & 214.2262 &  52.4111& 3.59$\times 10^{-15}$  & (1.78$\pm$0.07)$\times 10^{-15}$   &  (3.82 $\pm$ 0.34)$\times 10^{-16}$ &  0.2466 &  SF  \\
GROTH--37457 & 214.7951 &  53.2660& 4.38$\times 10^{-15}$ & (2.43$\pm$0.12)$\times 10^{-15}$   & $<$8.00$\times 10^{-16}$  &  0.2639 &  AGN \\
GROTH--31403 & 214.2910 &  53.0867& 2.25$\times 10^{-15}$ & (1.10$\pm$0.06)$\times 10^{-15}$   &  (2.99  $\pm$ 0.55)$\times 10^{-16}$  &  0.2672 &  SF  \\
GROTH--21404 & 215.1861 &  52.8351& 2.53$\times 10^{-15}$ & (2.06$\pm$0.06)$\times 10^{-15}$   &  (5.30 $\pm$  0.29)$\times 10^{-16}$  &  0.2518 &  SF  \\
GROTH--3525  & 214.7796 &  52.3522& 2.41$\times 10^{-15}$  & (1.30$\pm$0.11)$\times 10^{-15}$   & $<$1.79$\times 10^{-15}$  &  0.2649 &  AGN\tablenotemark{(d)} \\
GROTH--37380 & 215.1904 &  53.3248& 1.93$\times 10^{-15}$ & (1.32$\pm$0.08)$\times 10^{-15}$   &  (4.92 $\pm$ 0.38)$\times 10^{-16}$ &  0.2631 &  SF\\  
GROTH--12279 & 214.3008 &  52.5991& 2.08$\times 10^{-15}$ & (1.34$\pm$0.06)$\times 10^{-15}$   &  (5.02 $\pm$ 0.43)$\times 10^{-16}$ &  0.2612 &  SF  \\
GROTH--20285 & 215.1330 &  52.7994& 4.11$\times 10^{-15}$ & (1.41$\pm$0.03)$\times 10^{-15}$   &  (4.08 $\pm$ 0.28)$\times 10^{-16}$ &  0.2517 &  SF  \\
GROTH--7430  & 214.4311 &  52.4683& 7.67$\times 10^{-15}$  & (2.03$\pm$0.08)$\times 10^{-15}$   &  (6.85 $\pm$ 0.91)$\times 10^{-16}$ &  0.2079 &  SF  \\
NGPDWS--23690& 219.8446 &  35.3075& 5.88$\times 10^{-15}$ & (3.86$\pm$0.10)$\times 10^{-15}$    &  (9.95 $\pm$ 0.63)$\times 10^{-16}$ &  0.2477 &  SF  \\
NGPDWS--10002& 219.0922 &  34.9421& 4.33$\times 10^{-15}$ & (2.06$\pm$0.06)$\times 10^{-15}$    &  (7.52 $\pm$ 0.44)$\times 10^{-16}$ &  0.2685 &  SF  \\
NGPDWS--33782& 219.5770 &  35.6305& 6.05$\times 10^{-15}$ & (1.80$\pm$0.08)$\times 10^{-15}$   &  (5.45 $\pm$  0.39)$\times 10^{-16}$ &  0.2618 &  SF  \\
NGPDWS--11927& 219.1004 &  34.9935& 4.34$\times 10^{-15}$ & (7.09$\pm$0.48)$\times 10^{-16}$   & ... &  0.2145 &  SF\tablenotemark{(d)}  \\
NGPDWS--23216& 218.6954 &  35.2844& 4.09$\times 10^{-15}$ & (3.66$\pm$0.40)$\times 10^{-15}$   & $<$5.74$\times 10^{-15}$ &  0.1902 &  SF  \\
NGPDWS--6731 & 219.1529 &  34.8428&  1.01$\times 10^{-14}$ & (4.41$\pm$0.11)$\times 10^{-15}$  &  (9.61 $\pm$ 0.47)$\times 10^{-16}$ &  0.2796 &  AGN  \\
NGPDWS--28521& 219.0262 &  35.4586& 4.05$\times 10^{-15}$ & (1.52$\pm$0.08)$\times 10^{-15}$   &  (4.07 $\pm$0.52)$\times 10^{-16}$ &  0.2493 &  SF  \\
NGPDWS--35813& 219.0558 &  35.7291& 3.70$\times 10^{-15}$ & (7.41$\pm$0.47)$\times 10^{-16}$   &  (2.39 $\pm$0.30)$\times 10^{-16}$&  0.2596 &  SF  \\
NGPDWS--32840& 219.2433 &  35.5977& 2.32$\times 10^{-15}$ & (3.83$\pm$0.15)$\times 10^{-15}$  & $<$5.69$\times 10^{-16}$ &  0.2058 &  AGN\tablenotemark{(d)} \\
SIRTFFL--14450& 259.2110&  59.9642&   2.22$\times 10^{-14}$ & (2.14$\pm$0.02)$\times 10^{-14}$    &  (6.10 $\pm$0.14)$\times 10^{-15}$ &  0.1815 &  SF \\ 
SIRTFFL--14085& 258.1492&  59.9468&   2.87$\times 10^{-15}$ & (3.21$\pm$0.07)$\times 10^{-15}$    &  (9.83 $\pm$0.80)$\times 10^{-16}$ &  0.2203 &  SF  \\
SIRTFFL--10895& 258.5918&  59.8333&   6.18$\times 10^{-15}$ & (1.23$\pm$0.05)$\times 10^{-15}$  & ... &  0.2297 &  SF\tablenotemark{(d)}  \\
\hline
\tablenotetext{(a)}{      \hspace{1mm}\lya\ flux from Deharveng et al. (2008).}
\tablenotetext{(b)}{Redshift computed from the optical emission lines.}
\tablenotetext{(c)}{Class: Galaxy classification based on the optical
  emission lines. SF$=$star--forming galaxy, AGN$=$ active nucleus. }
\tablenotetext{(d)}{Classification in SF or AGN based on the \nii/\ha\
  ratio. GROTH--23096 was classified as an AGN based on the width of
  the \oiii\ emission lines. }
\end{tabular}                       
\end{table}

\end{document}